\documentstyle[preprint,aps,epsf]{revtex}

\begin{document}
\draft

\title{ARPES in the normal state of the 
cuprates: comparing the marginal Fermi liquid and spin fluctuation scenarios}
\author{R. Haslinger$^1$, Ar. Abanov$^1$ and A. Chubukov$^2$}
\address{
  $^1$ Los Alamos National Lab, Los Alamos, NM 87545\\
  $^2$ Department of Physics, University of Wisconsin, Madison, WI 53706
}

\maketitle

\begin{abstract}
We address the issue whether ARPES measurements of the 
spectral function $A_k (\omega)$ near the Fermi surface 
in the normal state of near optimally doped cuprates can distinguish between
 the marginal Fermi liquid scenario and the spin-fluctuation scenario. 
We argue that the data for momenta near the Fermi surface are equally well 
described by both theories, but this agreement is nearly meaningless 
as in both cases one has to add  to $\Sigma^{\prime \prime} (\omega)$ a large 
constant of yet unknown origin. We show that the data can be well fitted by keeping only 
this constant term in the self-energy. To distinguish
 between the two scenarios, one has to analyze the data away from the Fermi surface, 
 when the intrinsic  piece in $\Sigma (\omega)$ becomes dominant. 
\end{abstract}


The observed  discrepancies between the
 normal state properties of the cuprates and the 
predictions  of Fermi liquid theory  continue to attract a lot of
attention from the condensed-matter community. 
In a Fermi liquid, the quasiparticle damping near the Fermi surface behaves
 as $\omega^2$ or $(\pi T)^2$, whichever is larger. This gives rise to   
the $\omega^2$ ($T^2$) behavior of the width of the peak in the 
quasiparticle spectral function $A_k (\Omega) = 
(1/\pi) Im G(k, \Omega)$  measured in 
 the angle-resolved photoemission (ARPES) experiments. 

The data on optimally doped cuprates show several features which are in 
 disagreement with Fermi liquid theory: $\Sigma^{\prime \prime} 
(\omega, T)$ is large and does not display a Fermi liquid form
 for all $T > T_c$ and for all frequencies. Instead, over
 a wide range of $\omega$ and $T$, $\Sigma^{\prime \prime} (\omega, T)$ resembles 
 a linear function of both arguments~\cite{valla,kaminski_1,johnson1,bogdanov1}.
As a result $\Sigma^{\prime \prime} (\omega, T)$ does not
 become progressively smaller than $\omega$ with decreasing $\omega$, and 
the spectral function does not undergo a 
 drastic sharpening at the smallest energies, as it does in a Fermi liquid. 
Whether this effect is quantitative or qualitative is still a matter of debate
 as in the normal state experiments, $T$ remains finite and larger than $T_c$. The 
 data for overdoped materials do show a stronger sharpening of the spectral 
 function near $k_F$ indicating that these materials are likely Fermi liquids~\cite{over}. 
From this perspective, the unusual behavior of optimally doped cuprates 
 indicates that the upper energy cutoff for the Fermi liquid behavior
 rapidly reduces down from the Fermi energy. This is a typical situation for 
 system near a quantum phase transition.

In this communication, we compare two scenarios for the non-Fermi liquid behavior, 
both are based on the idea of quantum criticality. The first is 
the marginal Fermi liquid (MFL) scenario developed in Ref~\cite{mfl}. 
This scenario
 assumes that there exists a quantum-critical point (QCP) separating the 
Fermi liquid and the pseudogap regimes (Fig. 1a). Superconductivity
develops in the shaded region about the QCP.  In the quantum-critical regime, the
 system is assumed to display a marginal Fermi liquid behavior with 
\begin{equation}
\Sigma^{mfl}_{\mathbf k} (\omega) = -\lambda_{mfl} 
[ \omega \log (\frac{\omega}{\omega_0}) - i \frac{\pi}{2} \sqrt{\omega^2 + \pi^2T^2}]
\label{mfl}
\end{equation}
where  $\lambda_{mfl}$ and $\omega_0$ are the two 
adjustable parameters. To satisfy Kramers-Kronig relations, 
$\omega_0$ should possess a nonsingular dependence on $(\pi T/\omega)^2$, but this dependence is usually neglected.
 The self energy is defined such that 
$G({\mathbf k},\omega) = 1/(\omega - \epsilon_{\bf k} + \Sigma({\bf k},\omega) )$.
We plot  $\Sigma^{mfl} (\omega)$ in Fig. 1b and c.
 This marginal behavior naturally emerges near the critical points in $3D$, however 
 obtaining this behavior in 2D systems is still problematic. To do so, 
it is possible that special requirements such as, e.g., 
 singular momentum dependence of the low-energy bosonic mode which 
mediates the  fermionic interaction~\cite{khvesh} must be satisfied.

Another scenario is based on the idea that the dominant interaction in the cuprates 
is between fermions and their low-energy collective spin excitations.
This physics is described by the spin-fermion (SF) model.
~\cite{chub,Abanov01}
In this scenario, the
  non-Fermi liquid behavior in the normal state is also associated 
 with the  closeness to a critical point, but this point now separates paramagnetic 
 and antiferromagnetically ordered phases. 
 The pseudogap phase emerges in the SF model as a dome around this critical point, and 
 optimal doping roughly corresponds to a situation when
 this dome intersects the crossover line separating Fermi liquid and quantum-critical 
 regions (see Fig. 1d). Superconductivity results in the shaded region where
 the pseudogap and Fermi liquid regimes intersect, and the normal state 
 non-Fermi liquid behavior occurs in the quantum critical regime that near optimal 
 doping stretches down to almost $T_c$.
 The quantum-critical behavior in the SF model 
has been analyzed in detail. \cite{Abanov01} The model has two parameters; $\omega_{sf}$ which is the typical 
 frequency for the relaxational spin dynamics and also turns out to be the upper cutoff of the 
Fermi liquid behavior, and the dimensionless coupling $\lambda_{sf}$. 
 Both  $\lambda_{sf}$ and $\omega_{sf}^{-1}$ 
increase as the system approaches a magnetic instability,  but the product 
${\bar \omega} = 4 \lambda_{sf}^2 \omega_{sf}$ remains finite and serves as the 
 upper cutoff for the quantum-critical behavior. The separation into Fermi liquid 
 and quantum critical regions makes sense when $\lambda_{sf} \geq 1$ or equivalently
 when $\omega_{sf} \ll {\bar \omega}$. 
In general,  $\lambda_{sf}$ and $\omega_{sf}$  (but not ${\bar \omega}$) 
depend on the momenta along the Fermi surface, but this dependence is rather 
mild as we will show.

The microscopic nature of the SF model implies that the self-energy is totally 
determined and has to be obtained in  explicit calculations.
At $T=0$, the Eliashberg-type calculations yield    
\begin{equation}
\Sigma ^{sf}_{\mathbf k}\left( \omega \right) =\lambda_{sf} ~\frac{2\omega%
}{1+\sqrt{1 -i\frac{\left| \omega\right| }{\omega _{\mathrm{sf}}}}}
\label{sf}
\end{equation}
At small frequencies, $\omega \ll \omega _{\mathrm{sf}}$,
$\Sigma^{sf}_{\mathbf k}\left( \omega \right) =\lambda_{sf} \left(
\omega +i\omega \left| \omega \right|/4\omega _{\mathrm{sf}}\right)$
and the system displays a  Fermi liquid behavior.
At frequencies above $8 -10 \omega _{\mathrm{sf}}$,
 the fermionic self-energy displays a quantum-critical behavior, in which
 ${\bar \omega}$ is the only energy scale:
$\Sigma^{sf}_{\mathbf k}\left( \omega \right) =
e^{i\pi/4} \mathrm{sign} \omega ~\left( \bar{\omega }|\omega |\right) ^{1/2}$.
At intermediate frequencies, $\Sigma^{sf}_{\mathbf k}\left( \omega \right)$ 
interpolates between the two limits. Although there is no intermediate 
asymptotic (i..e., the crossover happens at energies $O(\omega_{sf})$), 
 the intermediate regime is rather wide, and in between $\omega_{sf}/2$ and 
at least $6-8 \omega_{sf}$,  $\Sigma_{\mathbf{k}}^{sf}\left( \omega \right)$  
resembles a linear function of  $\omega $. 
The finite temperature result is more complicated \cite{Abanov01} but
posesses the same basic characteristics as the zero temperature result.
We plot $\Sigma_{\mathbf{k}}^{sf}
\left( \omega \right)$ at $T=100K$ in Fig. 1d.

 In light of the strong simliarities between the MFL and SF fermionic
 self energies, it is important to ask which experiments allow us to
 distinguish between the two scenarios.
 We focus in this communication on  {\it energy distribution
curve} (EDC) ARPES experiments performed for momenta near the Fermi surface.
 Abrahams and Varma~\cite{av} have argued that such ARPES experiments help put strong constraints
 on the form of the pairing interaction and can rule out many of the scenarios  for the 
 non-Fermi liquid behavior in the normal state. In particular, they 
 argued (i) that  the MFL form of $\Sigma (\omega)$ nicely fits the experimental data, 
 and (ii) the data along the Fermi surface can be fitted assuming 
 that the MFL coupling $\lambda_{mfl}$  is almost independent of the position on
 the Fermi surface. They argued that this  
 rules out a magnetic scenario since the bosonic interaction is peaked at $Q=(\pi,\pi)$ 
 and hence the 
 coupling constant should be momentum dependent with the maximum at hot spots
 (points at the Fermi surface separated by $Q$) in contradiction with (ii).
  We, on the contrary, demonstrated previously \cite{rob1} that the 
 photoemission data near $(0,\pi)$ and along the zone diagonal can be well 
 fitted by the spin-fluctuation self-energy, Eq. (\ref{sf}), with  $\lambda_{sf}$
 {\it varying} along the Fermi surface.

We re-examine this issue using recent photoemission results 
from the Argonne group.~\cite{kaminski} 
We will argue that EDC data for $k \approx k_F$
 can be reasonably well fitted by both the MFL and SF theories, but the way in which
 the fitting
 works poses a serious question for both scenarios. Namely, in both fits, one has
   to add to the self-energy a large imaginary constant 
 $i \gamma ({\vec k}_F)$  of yet unknown origin. We show that 
 the fits with $\Sigma_k (\omega) = i \gamma ({\vec k}_F)$ with {\it no frequency 
 dependent terms added} turn out to be of the same quality as the fits with 
 $\Sigma^{full}_k (\omega) = \Sigma_k (\omega) + i \gamma$ where   $\Sigma_k (\omega)$ 
 has either MFL or SF form. From this perspective, the ARPES measurements near the 
 Fermi surface do not allow the ruling out of either scenario.
The situation may be different at large deviations from $k_F$ when the maximum in the 
photoemission intensity moves to high frequencies, and $\Sigma_k (\omega)$
 (which increases with $\omega$) becomes larger than $\gamma$. In this case,
 one could possibly distinguish between MFL and SF scenarios by examining 
 both the frequency dependence of $\Sigma^{\prime \prime} (\omega)$ and the
 sign and magnitude of $\Sigma^{\prime} (\omega)$.

Before we proceed with the fitting, we briefly review the experimental 
setup~\cite{valla,kaminski_1,bogdanov1,norman,shen,dessau,millis}.
Two types of ARPES experiments are currently available. The EDC 
(energy distribution curve) experiments  measure fermionic 
$I_k(\omega) = A_k(\omega) n_F(\omega)$ as a function of 
frequency at a given $k$. The MDC (momentum distribution curve) 
 experiments measure $I_k (\omega)$ 
as a function of $k$ transverse to the Fermi surface 
at a given frequency. In general, near the Fermi surface,
\begin{equation}
A_k (\omega) = \frac{1}{\pi} \frac{\Sigma^{\prime \prime}_k (\omega)}{(\omega + 
\Sigma^\prime_k(\omega) - v_k (k - k_F))^2 + (\Sigma^{\prime \prime}_k (\omega))^2}
\label{1}
\end{equation}

There are both experimental and theoretical indications that in the cuprates, 
the fermionic self-energy is almost independent of the momentum component 
transverse to the Fermi surface although it strongly depends on frequency, 
and also possesses some dependence on the momentum along the Fermi surface.
 In this situation, the self-energy  in the MDC measurements remains 
almost a constant, and 
 the photoemission intensity evolves with $k$ as 
\begin{equation}
I^{MDC}_k (\omega) \propto \frac{1}{(v_F (k - k^* (\omega)))^2 + 
(\Sigma^{\prime \prime})^2}
\label{2}
\end{equation}   

 where $k^* (\omega) = k_F - (\omega + \Sigma^\prime (\omega))/v_F$.
We see that $I_k$ has a Lorentzian form with the maximum at $k = k^*$ 
 and FWHM $|k-k^*| = \Sigma^{\prime \prime} (\omega)/v_F$. 
In other words, the MDC measurements allow one to directly infer the frequency
 dependence of the fermionic self-energy.  
If $v_F$ were known,  the MDC measurements would also 
yield  the magnitude of $\Sigma^{\prime \prime} (\omega)$. However to obtain $v_F$
 is itself a problem as the MDC dispersion extracted from the peak position
 at low enough $\omega$ is $\omega (k) = v^*_F (k - k_F)$ where $v^*_F = v_F 
(1 + \partial \Sigma^{\prime} (\omega)/\partial \omega)$, hence measuring the 
dispersion at low frequencies 
one obtains a renormalized  $v^*_F$, not a bare $v_F$~\cite{kaminski_1}. 
One could hope 
 to extract $v_F$ from the dispersion at  higher
 frequencies, where $\omega >  \Sigma^{\prime} (\omega)$, 
and $v^*_F$ approaches $v_F$. However, it is not quite clear whether 
 $ \Sigma^{\prime} (\omega)$ can be fully neglected at energies where one can 
 restrict with the linearized dispersion near the Fermi surface.

In the EDC experiments, one measures at small enough frequencies
\begin{equation}
I^{EDC}_k (\omega) \propto \frac{\Sigma^{\prime \prime} (\omega)}{(\omega - 
\omega (k))^2 + (\Sigma^{\prime \prime} (\omega)/v^*_F)^2}
\label{3} 
\end{equation}
In a Fermi liquid $\Sigma^{\prime \prime} (\omega)$ is small and near 
$\omega = \omega (k)$ can be approximated by a constant - 
its value at $\omega = \omega (k)$. $I^{EDC}_k (\omega)$ then has a Lorentzian 
form with FWHM $\Sigma^{\prime \prime} (\omega)/v^*_F$. As $v^*_F$ is directly 
measured from the dispersion, one could straightforwardly 
extract  $\Sigma^{\prime \prime} (\omega)$ from the data. In the cuprates,
however, the situation is more complex as 
(i) typical frequencies are not small, and hence $v_F^*$ itself depends on 
frequency, and (ii) $\Sigma^{\prime \prime} (\omega)$ is not small, and its 
frequency dependence matters. Both effects give rise to a non-Lorentzian form of 
$I^{EDC}_k (\omega)$, and the extraction of 
$\Sigma^{\prime \prime} (\omega)$ becomes problematic.  

In Figs \ref{fig2}, \ref{fig4} and \ref{fig5} we present the data for 
 $I_k(\omega)$ obtained from EDC experiments
by Kaminski {\it et al} \cite{kaminski} taken at $100K$ from near optimally 
doped Bi2212 with a $T_c$ of $85K$.   We have chosen lineshapes for two points near the 
Fermi surface, one is close to the zone diagonal ($k_y=0.54 \pi/a$)
, and another near a hot spot ($k_y=0.81 \pi/a$).  We see that the
form of $I^{EDC}_k (\omega)$ is clearly different from a Lorentzian, even when
the Fermi function is taken into account. In this 
situation, it is more reasonable to fit the whole  $I^{EDC}_k (\omega)$ rather 
than the FWHM. This is what we will do.

\begin{figure}[tb]
\epsfxsize \columnwidth
\epsffile{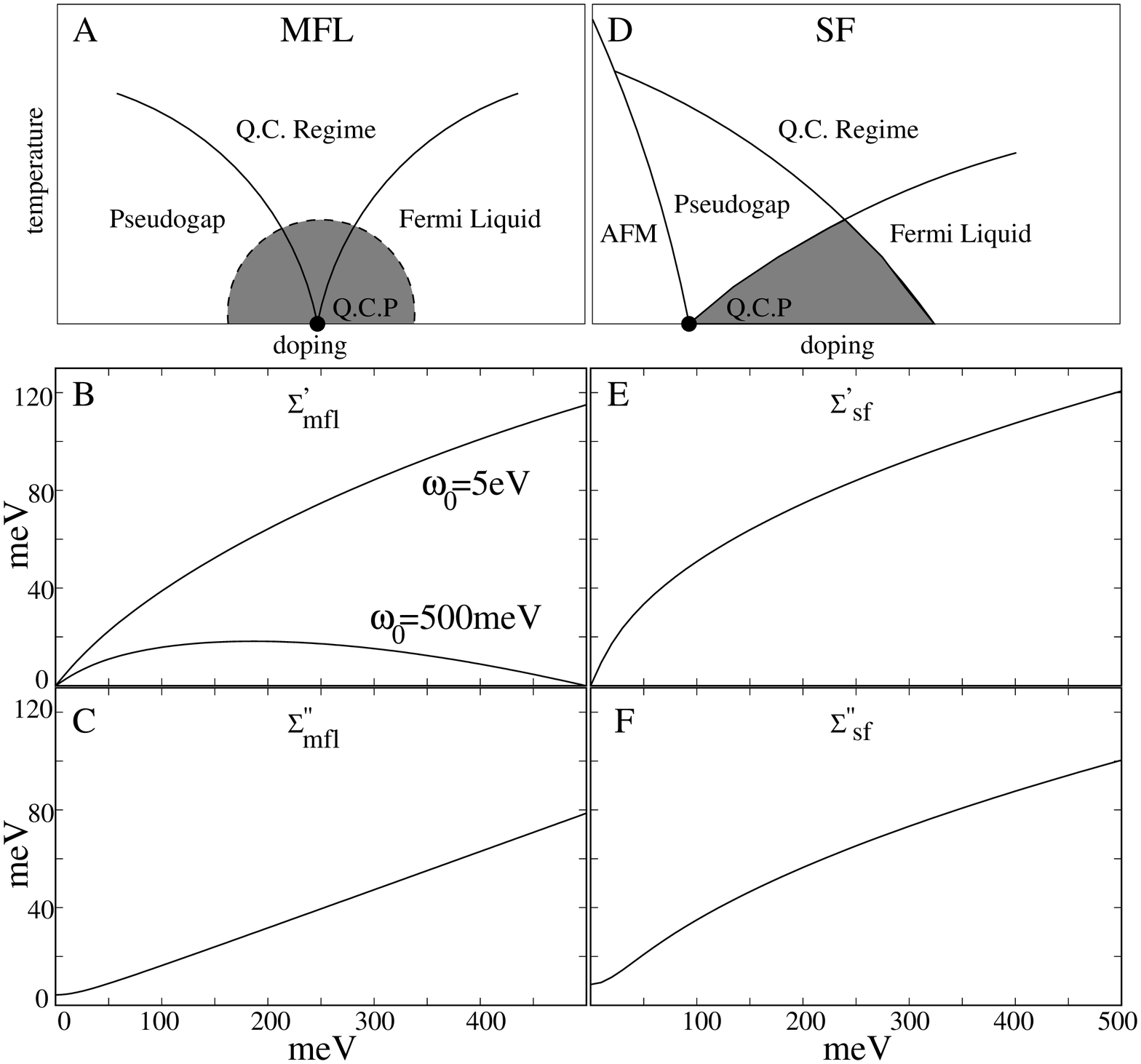}
\caption{Comparison of the marginal fermi liquid and spin fermion scenarios.
Figures a and d show the proposed phase diagrams for the MFL (a) and SF (d)
pictures.  Descriptions of the phase diagrams are in the text. 
Figures b,c and e,f compare the MFL and SF self energies at $T=100K$.
Figures b and c show the real and imaginary parts of the MFL self energy respectively.
$\lambda_{mfl}=0.1$ in both figures.  In figure b, $\Sigma'$ is given for
both $\omega_0=500meV$ and $\omega_0=5eV$.  Figures e and f show the real
and imaginary self energies for the spin fermion model. $\lambda_{sf}=1.0$
and $\omega_{sf}=15meV$.  $\Sigma^{\prime \prime}$ 
is seen to be roughly linear over 
a wide frequency range.}
\label{fig1}
\end{figure}

\begin{figure}
\epsfxsize \columnwidth
\epsffile{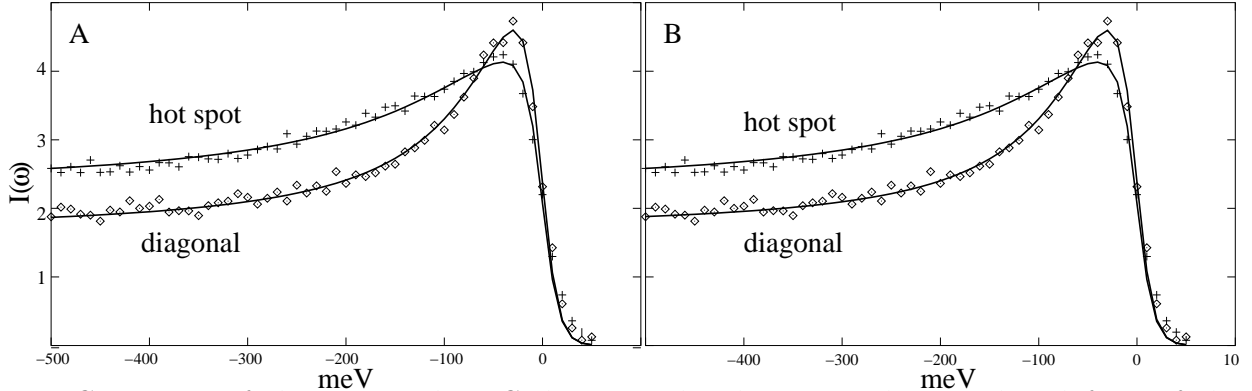}
\caption{Fits of the measured EDC dispersion by the marginal Fermi liquid form of 
the self-energy. The intensity $I(\omega)$ is in arbitary units.
The two fits have almost the same coupling $\lambda_{mfl}$ but 
different $\omega_0$ and different $\gamma$. 
In Fig. a, $\omega_0 = 500 meV$ in which case the real part of the self-energy 
is small for the relevant frequencies. 
In Fig. b, $\omega_0 = 5 eV$, and 
$\Sigma^{\prime}$ is relevant. Both fits work very well.
 The parameters are presented in the text.}
\label{fig2}
\end{figure}

In Fig.\ref{fig2} we present the 
theoretical fits using the MFL form of the self-energy. 
The fit in Fig. \ref{fig2}a is obtained using $\omega_0 = 500 meV$. 
 For this $\omega_0$, $\Sigma^{\prime} (\omega)$ is small at
frequencies of few hundred meV (see Fig. \ref{fig1}) 
and to first approximation can be neglected.
The fit in Fig.  \ref{fig2}b is obtained using a deliberately large 
$\omega_0 =5 eV$, in which case the real part of the self-energy is strong.
We see that both fits are rather good.
The fitting parameters in Fig \ref{fig2} a (b) 
are $\lambda_{mfl} =0.17 (0.17)$, $\gamma =100 (133) meV$ and $\epsilon_k = -26 (-30) meV$ 
for $k$ along the zone diagonal, and $\lambda_{mfl} = 0.15 (0.15),
\gamma = 171 (225) meV$, and $\epsilon_k = -43 (-47) meV$
for $k$ close to a hot spot.
To account for the flattening of the measured $I^{EDC}_k (\omega)$ at the highest 
frequencies, we also had to add a background contribution to $I^{EDC}_k (\omega)$ 
which we have chosen, following Ref. \cite{av} to be just 
proportional to the Fermi function:
$I^{EDC}_k (\omega) \rightarrow I^{EDC}_k (\omega) + \beta n_F (\omega)$. 
We found $\beta = 1.57 (1.64)$ along zone diagonal and $\beta = 2.22 (2.26)$ 
near the hot spots.

\begin{figure}
\begin{center}
\epsfxsize .5\columnwidth
\epsffile{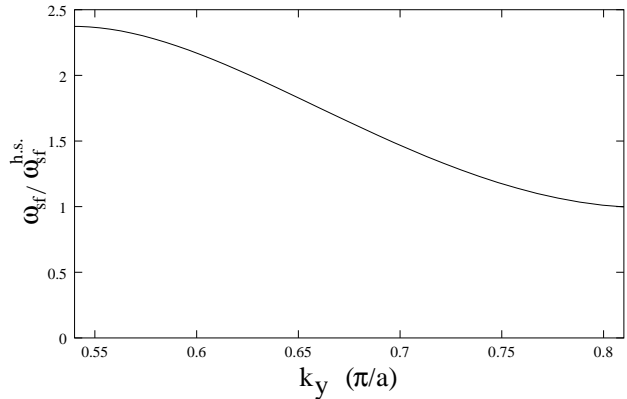}
\caption{The behavior of the spin-fluctuation frequency $\omega_{sf}$ along
the Fermi surface.  This frequency is the upper cutoff for the Fermi liquid
behavior near an antiferromagnetic instability. 
We used 
 $\epsilon(k_x,k_y) = -2t(\cos k_x + \cos k_y ) + 4t' \cos k_x \cos k_y - \mu$
 with $t=250 meV$, $t'=-0.36t$, and $\mu=-1.1t$ chosen as being
representative of optimal doping.  
The variation of $\omega_{sf}$  at a deviation from hot spots 
is determined by a competition between the increase 
 due to a reduction of the magnetically mediated interaction,
 and the decrease due to the ``nesting tendency'' - 
the increase  towards $\pi$ of the 
angle between the Fermi velocities at ${\bf k}_F$ and ${\bf k}_F +{\bf Q}$.}
\label{fig3}
\end{center}
\end{figure}

\begin{figure}
\epsfxsize \columnwidth
\epsffile{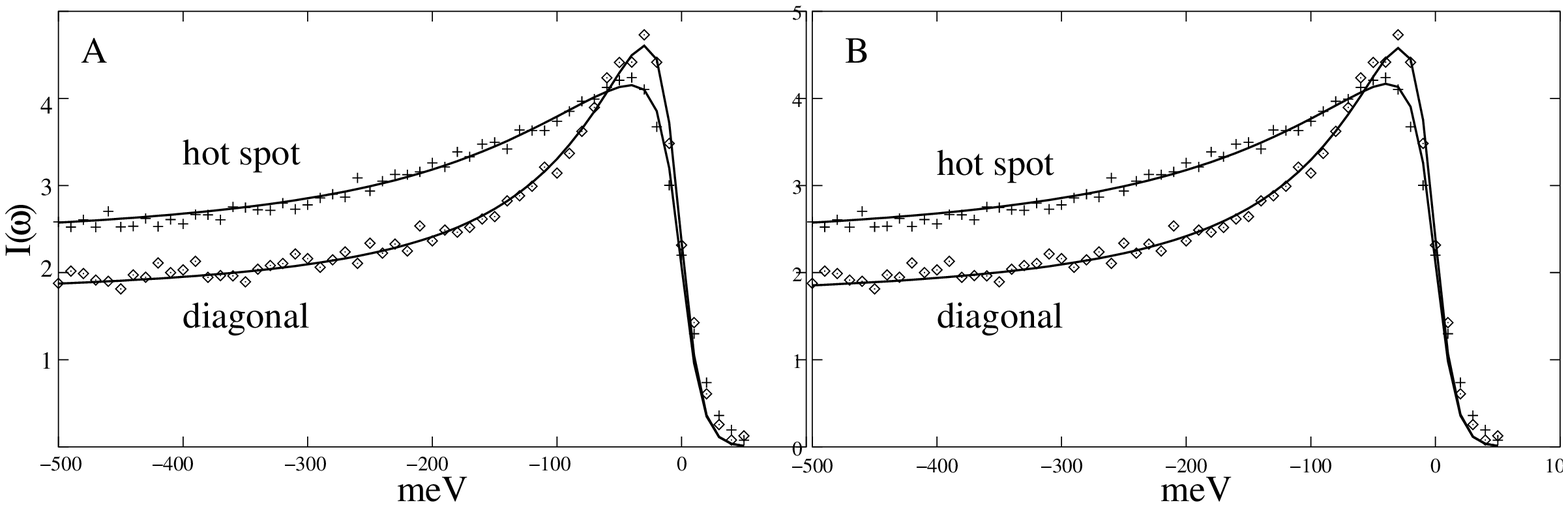}
\caption{Fits of the measured EDC dispersion by the
spin-fermion model with $\omega_{sf} = 15 meV$. 
(a) a fit using  momentum independent coupling 
$\lambda_{sf} (k)=0.84$ for both the diagonal and the hot spot,
 (b) a fit using  momentum dependent coupling  
 $\lambda_{sf} = 1.0$ along the diagonal and $\lambda_{sf} =1.5$ 
near the hot spots. Other parameters are presented in the text.}
\label{fig4}
\end{figure}

\begin{figure}
\begin{center}
\epsfxsize .5\columnwidth
\epsffile{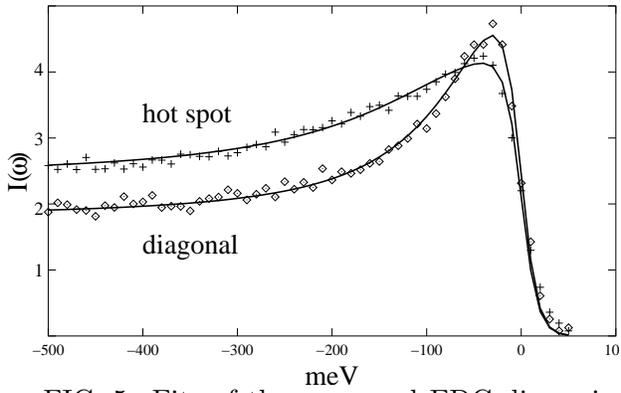}
\caption{Fits of the measured EDC dispersion by the
 self-energy $\Sigma (\omega) = i \gamma (k)$ 
 with no extra MFL or SF frequency dependence. The
 parameters are presented in the text.}
\label{fig5}
\end{center}
\end{figure}

We next consider the fits by the SF model. As stated previously,  
the two input parameters in the model, $\omega_{sf} $ and $\lambda_{sf}$ generally 
depend on the momentum component along the Fermi surface. 
Near the hot spots, this dependence is universal and is given by 
$\lambda_{sf} (k)= \lambda_{sf} /Z^{1/2}_k, ~\omega_{sf} (k) = 
\omega_{sf} Z_k$ where 
 $Z_k = 1+(\delta k\xi )^{2}$, and  $\xi$ and $\delta k$ are the spin correlation length and the momentum
deviation from a hot spot along the Fermi surface, respectively~\cite{Abanov01}. 
The largest $\delta k\xi$ is for $\mathbf{k}$ along the zone diagonal. 
At optimal doping, ARPES measurements of the Fermi surface
 yield $\delta k^{\mathrm{\max }}\sim 0.2\pi /a\approx 0.6/a$~\cite{valla,kaminski_1}.  The correlation length extracted 
  from the NMR measurements~\cite{nmr}  is $\xi \sim 1-2 a$. Then 
$\lambda_{sf} $ is reduced by at most a factor of  $2$ as one moves
from hot spots to zone diagonal. For $\omega_{sf}$, the use of the expansion 
formula near hot spots would yield a bigger increase, but its actual variation 
is much smaller as the theoretical $\omega _{\mathrm{sf}%
}\propto \sin \phi _{0}$, where $\phi _{0}$ is the angle between Fermi
velocities at $\mathbf{k}$ and $\mathbf{k}+\mathbf{Q}$, which  tends to 
$\pi$ as $\mathbf{k}$ approaches the zone diagonal. \cite{chub}
 In Fig. \ref{fig3} 
we present the results of a computation of $\omega_{sf} (k)$ along the Fermi 
surface which includes both of the above effects.
The increase is only a factor of 2.4, consistent with the 
fact that $\omega_{sf} (k_{diag})$ extracted from fitting $v^*_F$ by the 
SF self-energy yields $\omega_{sf} (k_{diag}) \sim 20-25 meV$
~\cite{Abanov01} 
while the fits to NMR yield $\omega_{sf} \sim 15 meV$ for fermions near hot 
spots~\cite{nmr}. For simplicity, in the SF fits,
 we will consider  $\omega_{sf}$ to be independent of $k$  and 
set $\omega_{sf} =15 meV$ in accordance with NMR. 
We, however,  have verified that variations
in $\omega_{sf}$ have little effect on ARPES lineshapes.

 The fits  using the SF form of the self-energy are presented 
 in Fig~\ref{fig4}. 
The two fits are obtained using momentum independent and 
 momentum-dependent  coupling
 $\lambda_{sf} (k)$, respectively. We see that the fits are of the same quality as those in Fig.~\ref{fig2}. The fitting parameters in Fig \ref{fig5} a (b) 
are $\lambda_{sf} = 0.84 (1.0)$, $\gamma =110 (114) meV$, 
 $\beta = 1.68 (1.63)$ and $\epsilon_k = -32 (-30) meV$ 
for $k$ along the zone diagonal, and $\lambda_{sf} = 0.84 (1.5),
 \gamma = 180 (193) meV$, $\beta = 2.31 (2.28)$ 
and $\epsilon_k = -52 (-45) meV$
 for $k$ close to a hot spot. We verified that making $\lambda_{sf}$ a bit larger 
 or allowing it to vary more over the Fermi surface does not change the accuracy of the fits.

Since the frequency dependence of the self energy appears to make
little difference in the quality of the fits at the Fermi surface,
we now attempt to fit the data by 
$\Sigma (\omega) = i \gamma (k)$ 
{\it without} extra MFL or SF frequency dependence. 
The results of these fits are presented in Fig.\ref{fig5}.
The fitting parameters are
$\gamma = 94meV$, 
$\beta = 1.8$ and $\epsilon_k = -4.4 meV$ 
for $k$ along the zone diagonal, and $
\gamma = 160 meV$, $\beta = 2.4$ 
and $\epsilon_k = -12 meV$
for $k$ close to a hot spot. $\epsilon_k$ is much smaller than in 
the other two fits, due to the lack of a real part of the self energy
in these fits.
 
All three fitting procedures work surprisingly well.
The facts that the data are not obtained exactly at the Fermi surface (i.e., 
$\epsilon_k \neq 0$) and that one has to add the background do not differentiate 
between the fits and hence are not that relevant, particularly as 
$\epsilon_k \sim 50 meV$ corresponds to a very small 
$|k-k_F| \sim \epsilon_k/v_F \sim 0.02 (\pi/a)$. 
The two physically relevant parameters are the coupling constant $\lambda$ 
and the extra $\gamma (k)$. 
By adjusting $\gamma (k)$ one can fit the data quite well by both 
MFL and SF forms, and also by just $\Sigma^{\prime \prime} = i \gamma (k)$.
This indicates that {\it one cannot differentiate between theoretical scenarios 
by analyzing EDC data obtained near the Fermi surface}. 
It is possible that the intrinsic piece of the self energy may be extracted
by fitting either EDC or MDC at deviations from the Fermi surface of at least a
few hundred meV.  Finally, the 
  large $\gamma (k)$ in the fits poses a problem for both MFL and SF scenarios. The origin of the  
$\gamma (k)$ is currently unknown and its understanding is clearly called for.

\acknowledgments
It is our pleasure to thank E. Abrahams, D. Basov, G. Blumberg, J.C. Campuzano
P. Johnson, A. Kaminski, M. Norman, D. Pines, J. Schmalian and C. Varma
 for useful discussions. We are also 
thankful to A. Kaminskii and J.C. Campuzano 
 for sending us the unpublished experimental data.
 The research was supported by NSF DMR-9979749 (A. Ch.) by DR Project 200153, 
 and by the Department
 of Energy, under contract W-7405-ENG-36.
 (R.H. and A.Ab.)

\end{document}